\documentclass[%
 reprint,
 amsmath,amssymb,
 aps,
]{revtex4-2}

\usepackage{graphicx}
\usepackage{dcolumn}
\usepackage{bm}

\usepackage[utf8]{inputenc} 
\usepackage[T1]{fontenc}  
\usepackage{mathptmx}     
\usepackage{etoolbox}     
\usepackage{setspace}     
\usepackage{etoolbox}     
\usepackage{xcolor,soul}  
\setstcolor{red}          
\usepackage[normalem]{ulem}
\usepackage{chngcntr}     

\begin{document}

\preprint{APS/123-QED}

\title{Granular flows over normally vibrated inclined bases}

\author{Prasad Sonar}
\affiliation{Department of Mechanical System Engineering, Tokyo University of Agriculture and Technology, 184-8588, Tokyo, Japan}
\affiliation{Department of Mechanical Engineering, Indian Institute of Technology Kanpur, 208016, Uttar Pradesh, India}
\email{prasadrsonar@gmail.com}

\author{Ashish Bhateja}
\affiliation{
 School of Mechanical Sciences, Indian Institute of Technology Goa, Ponda 403401, Goa, India
}%
\author{Ishan Sharma}
\affiliation{%
 Department of Mechanical Engineering, Indian Institute of Technology Kanpur, 208016, Uttar Pradesh, India
}%
\affiliation{Department of Space, Planetary \& Astronomical Sciences and Engineering, Indian Institute of Technology Kanpur, 208016, Uttar Pradesh, India}
\affiliation{Mechanics \& Applied Mathematics Group, Indian Institute of Technology Kanpur, 208016, Uttar Pradesh, India.}

\date{\today}

\begin{abstract}
We investigate granular flows over an inclined rigid base, which is vibrated externally in a direction normal to itself, through the discrete element method. We vary the base’s inclination angle $\theta$, vibration frequency $f$, and amplitude $A$ to study changes in the granular flow’s profile and the mass flow rate $Q$. We find that the flow velocity profiles for the vibrated bases are non-linear, unlike their fixed-base counterparts. Our study reveals that $Q$ may be maintained nearly constant in flows over vibrated bases utilizing appropriate combinations of $\theta$, $A$, and $f$.
At the same time, by vibrating the base depending on the inclination angle, we may increase the mass flow rate by $25$ to $100$ times, compared to the values for the fixed base. Finally, we characterize $Q$ in terms of a non-dimensional number $S$, which is the ratio of vibrational and gravitational energies.
\end{abstract}

\maketitle


\section{\label{sec:intro}Introduction}
%
%
Granular flows over inclined surfaces are common in industries and nature~\cite{daerr1999,hutter2003,goldhirsch2003}.
Such flows exhibit a variety of phenomena, including pattern formation~\citep{gray1997}, and segregation driven by gravity~\citep{gray2018}, temperature~\citep{neveu2022} and density-difference~\citep{tripathi2013}. Besides particle-level parameters, such as grain size, shape, friction, and inelasticity, boundary conditions play a crucial role in controlling the dynamics of inclined granular flows~\cite{pouliquen1999,silbert2001}. 
For instance, for boundary-driven flows, the base topography may drive the flow to ordered, disordered, and oscillating states~\citep{gray1999, silbert2002, schaefer2013, bharathraj2017, yang2022}. Such boundary effects are observed to penetrate deep into granular flows~\cite{bharathraj2017} and alter their characteristics.

A variety of phenomenology emerges when energy is injected into the system through base vibrations, e.g. development of various patterns like stripes, oscillations, squares, kinks and zippers~\citep{umbanhowar1996,das1998}. It is common to characterize vibrated granular systems using vibration frequency and amplitude~\cite{kumaran1998,umbanhowar1997,nowak1998,soto2004,bhateja2016,gaudel2016,watson2023}. 
Soto~\citep{soto2004} employed kinetic theory to study vibrated granular systems and found that the flow is controlled by the product of vibration amplitude and frequency, and not by these parameters alone.
Eshuis \textit{et al.}~\citep{eshuis2007} systematically performed experiments to study a shallow, vertically shaken quasi-two-dimensional horizontal granular bed and constructed a phase diagram in which the outcomes were distinguished by the presence of features such as bouncing bed, undulations, Leidenfrost effect, convection rolls, and granular gas as the fluidization increases. 
They characterized various phases of fluidized grains in terms of the number of layers, and dimensionless numbers $S = A^2 \omega^2 / g d$ and $\Gamma = A \omega^2 /g$, where $d$ is the grain diameter, $A$ is the amplitude of vibration, $g$ is the gravitational acceleration and  $\omega = 2 \pi f$, with $f$ being the vibration frequency. The phase diagram presented by the authors clearly distinguishes the regimes governed primarily by $S$ and $\Gamma$.
Pak \textit{et al.}~\cite{pak1994} in their experimental study employed $\Gamma > 1$ as a necessary condition for defining the sustained flow. Studies related to vertical discharge through horizontally vibrated hoppers also examine the dependence of depth-averaged mass flow rate $Q$ on $\Gamma$~\cite{hunt1999,kumar2020,wassgren2002}.
The number $S$ was found to regulate granular temperature, i.e. velocity fluctuations, and fracturing regimes in vibro-fluidized systems comprising coarser grains~\cite{bhateja2016} and fine cohesive powders~\citep{sonar2023}, respectively.

Gaudel and co-workers~\cite{gaudel2016} studied inclined granular flows using a \textit{transversely} vibrated base in their experiments, and obtained vibration-driven flows at inclinations where the gravity is insufficient to induce flow. A scaling law was proposed based on minimum $\Gamma$ required to activate the flow at low inclinations. This work was later supported by discrete element method (DEM) simulations of Gaudel \textit{et al.}~\cite{gaudel2019}, where the gravity-driven flows at large inclination angles were found to be consistent with the Bagnold rheology~\cite{bagnold1954}. However, the vibration-driven flows for inclinations below the critical angle were observed to deviate from Bagnold curve.
{Recently, d'Ambrosio \textit{et al.}~\cite{ambrosio2023} found experimentally that base vibrations not only help reduce inter-particle friction and control the flow, but also promote segregation of fine particles that decrease basal friction and, thus, act as a lubricant~\cite{ambrosio2023}.} It is worth remarking that such empirical observations help develop appropriate boundary conditions for theoretical treatments~\cite{warr1995,brey2000,soto2004} of vibrated granular systems.

Earlier, Zhu \textit{et al.}~\cite{zhu2020} numerically studied granular flows on inclined surfaces under \textit{normal} vibration to investigate earthquake-induced landslides, and also observed that mobility of flows is enhanced by low-frequency vibration.
{Niu \textit{et al.}~\cite{niu2024} recently discussed the effect of ground acceleration on thick granular flow under different gravity conditions. They found that vibration has a greater impact on the flow under low-gravity environments and low inclination angles.}

We note that the interplay of inclination angle and vibrational parameters is not well understood for grains flowing over vibrated inclined surfaces.
In this work, thus, we examine through DEM simulations granular flows over inclined bases vibrated in a direction normal to itself; see Fig.~\ref{fig:schematic}. 
We study relatively thin flows with depth of about ten particle diameters, which require a narrow range of vibrational amplitude and frequency due to their higher susceptibility to base vibrations.
For such systems, we characterize the flow rate in terms of $S$ and $\Gamma$, offering a way to control granular flow on vibrated bases.
To the best of our knowledge, such a correlation between the flow rate and these dimensionless numbers has not been explored for inclined granular flows.

The organization of this paper is as follows. We present details of numerical simulations in Sec.~\ref{sec:sim}. The flow profiles on the fixed and vibrated bases are discussed in Sec.~\ref{sec:obs}, including scaling of the flow rate. 
Finally, we conclude in Sec.~\ref{sec:con}.
%
%
\section{\label{sec:sim}Simulation methodology} 
%
%
\begin{figure}
\centering
\includegraphics[scale=0.42]{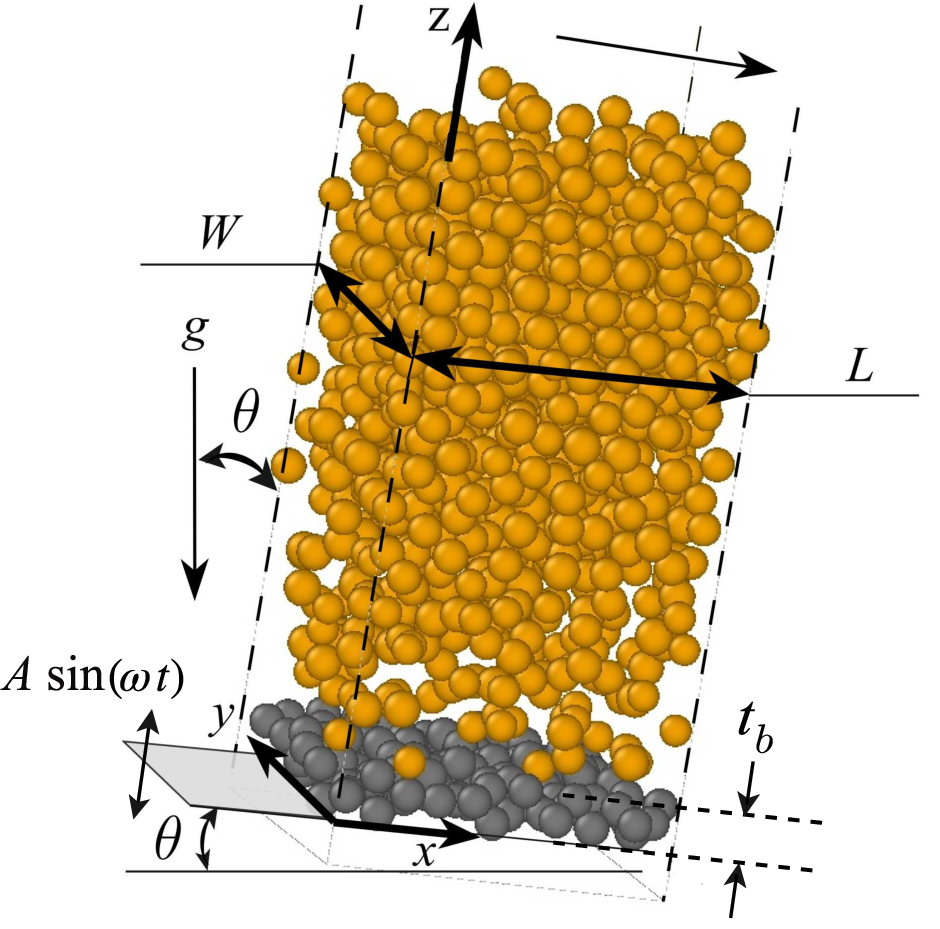}
\caption{A simulation snapshot presenting granular flow on a vibrated bumpy base.
The coordinate axes, directions of vibration and gravity are shown suitably. Gray-colored grains form the bumpy base, having a thickness of $t_b \approx 2d$.}
\label{fig:schematic}
\end{figure}
We study gravity-driven steady granular flows over an inclined bumpy base, which is vibrated sinusoidally with amplitude $A$ and frequency $f$ along the vertical ($z$) direction at $z_b(t)=A~\text{sin}(\omega \, t)$ as shown in Fig.~\ref{fig:schematic}. We consider a fixed coordinate system $x-y-z$, with $z =0$ corresponding to the bottom surface of the base when not vibrated. We perform computations based on the discrete element method (DEM)~\cite{cundall1979} utilizing the open-source package LAMMPS~\cite{plimpton2007}. Periodic boundary conditions are applied along the flow ($x$) and transverse ($y$) directions. The granular assembly consists of spherical grains of mean diameter $d$, with $\pm 10\%$ polydispersity in size. The grains are dry, cohesionless, inelastic, and rough. 
The bumpy base comprises randomly arranged frozen grains of diameter $d$. To create the base, grains are poured into a box under gravity. After they settle and come to rest, the grains located within $2d$ thickness around the mid-plane of the bed are extracted to create the bumpy base. The length $L$ and width $W$ of the base are $10d$ each. The static bed thickness is $h_0 \approx 10d$ corresponding to $N = 1000$ grains.

The interaction between grains is modeled through the $L3$ model of Silbert \textit{et al.}~\cite{silbert2001} and employs the simulation parameters as employed by them. This model utilizes the linear-spring dashpot force scheme to characterize collisions between grains and is commonly employed in computational studies based on the discrete element method~\cite{sun2006,reddy2010,tripathi2011,tripathi2013,mandal2017}.
The $L3$ model includes springs with stiffnesses $k_n$ and $k_t$, and dashpots with damping constants $\gamma_n$ and $\gamma_t$ for the normal ($n$) and tangential ($t$) directions, respectively. 
We note that Silbert \textit{et al.}~\cite{silbert2001} observed qualitative agreement between velocity profiles for the linear ($L3$) and non-linear (Hertzian, $H3$) models in their study of inclined granular flows. Later, 
Bharathraj and Kumaran studied the effect of base topography and roughness on the ordering and disordering of granular flows of varied thicknesses~\cite{bharathraj2017}. The linear and Hertzian models were used in their analysis, and a qualitative agreement was reported between both models for the velocity profiles.
We validate our simulations by comparing our findings with those of Silbert \textit{et al.}~\cite{silbert2001} for the $L3$ model. Our results differ slightly, the reason for which may be attributed to the difference in the characteristics of the bumpy base.

An important difference between Silbert \textit{et al.} and ours is that the base can vibrate. 
All results presented in this work are reported in terms of the non-dimensional quantities, normalized using an appropriate combination of the grain mass density $\rho$, diameter $d$, and acceleration due to gravity $g$. The quantity used for scaling time units is $\sqrt{d/g}$ and we set $f_g = \sqrt{g/d}$ as a scaling frequency. 
The parameters employed in the interaction model are $k_n = 2 \times 10^5$, $k_t = 2/7\, k_n$, $\gamma_n = 50$, $\gamma_t = 0$, where $k$ and $\gamma$ are the spring and damping constants, respectively, and the subscripts `$n$' and `$t$' denote the normal and tangential directions. The coefficients of restitution ($e$) and friction ($\mu$) are $e=0.88$ and $\mu = 0.5$. The time step ($\delta t$) of integration is $5 \times 10^{-5}$. The vibration amplitude and frequency are kept in the ranges $A=0.1d-0.5d$ and $f = 1 f_g - 5 f_g$, respectively.

The results reported here are obtained at steady state. The flow profiles are computed by dividing the simulation domain into horizontal bins parallel to the flow direction with thickness of $0.5d$. 
The mean flow fields are calculated by averaging the data over $4000$ snapshots. For the fixed base, snapshots are recorded at an interval of $500$ time steps. The averaging is carried out over 100 cycles for the vibrating scenario, with each cycle having $40$ time intervals. Accordingly, the time interval between snapshots varies with the vibration frequency.

The mean translational velocity $\hat{\upsilon}_x(z) = \upsilon_x(z)/\sqrt{gd}$ along $x$ in a bin centred at the {depth} $z$ is calculated from:
\begin{equation}
\hat{\upsilon}_x(z) =  \frac{1}{N_b} \sum_{i=1}^{N_b} c_{xi},
\label{vx}
\end{equation}
where $N_b$ is the number of grains in the bin and $c_{xi}$ is the component of the instantaneous velocity vector $\bm{c}_i$ of grain $i$ along $x$ direction. Similarly, the volume fraction $\phi$ in the bin is given by
\begin{equation}
\phi(z) =  \frac{1}{V_b} \sum_{i=1}^{N_b} V_i,
\label{Tx}
\end{equation}
where $V_i$ is the volume of grain $i$ having its center-of-mass located in the bin and $V_b$ is the bin volume. 
The mass flow rate $\hat{Q}_x(z) = Q_x(z)/ \rho d^3 \sqrt{g/d}$ in a bin may be expressed as
\begin{equation}
\hat{Q}_x(z) = \rho \ \phi(z) \ \hat{\upsilon}_x(z) \ A_b,
\label{qx}
\end{equation}
where $\rho$ is the mass density of grains and $A_b=L~W$. For simplicity, from now on, we represent non-dimensional quantities as $\upsilon_x$ and $Q_x$ instead of $\hat{\upsilon}_x$ and $\hat{Q}_x$, respectively. Finally, we compute the depth-averaged mass flow rate
\begin{equation}
Q = \int_{0}^{H} Q_x(z) dz,
\end{equation}
where $H$ is the thickness of the flowing granular bed at steady state. Note that $H$ varies with $A$, $f$ and the inclination angle $\theta$.
%
%
\section{Results}
\label{sec:obs}
\begin{figure}
\centering
\includegraphics[scale=0.33]{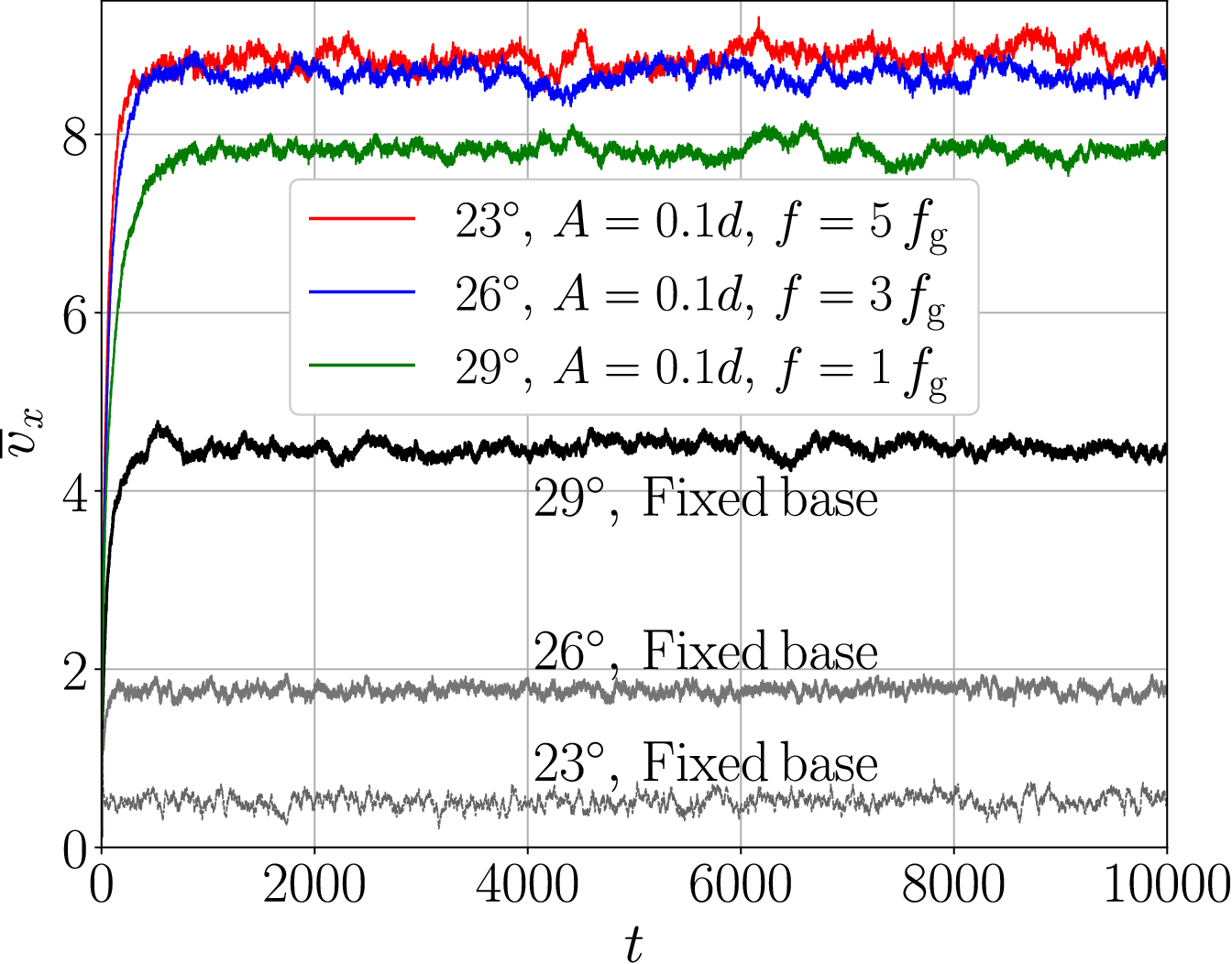}
\caption{Evolution of the average translational velocity $\overline{\upsilon}_x$ on the fixed and vibrated bases. The flow velocity is {scaled by $\sqrt{gd}$.} 
}
\label{fig:steady}
\end{figure}

Let us first find the range of $\theta$ wherein the flow achieves a steady state. 
To this end, we compute the average translational velocity $\overline{\upsilon}_x(t)$ of all grains along $x$ at time $t$, which is given as
\begin{equation}
\overline{\upsilon}_x(t) = \frac{1}{N}\sum_{i=1}^{N} c_{xi}.
\end{equation}
The temporal evolution of $\overline{\upsilon}_x$ for the fixed and vibrated bases for three inclination angles $\theta$ is shown in Fig.~\ref{fig:steady}. For the fixed base, the steady state occurs for all three values of $\theta$. The material ceases to flow for $\theta < 23^\circ$ and accelerates when $\theta$ grows above $29^{\circ}$. As expected, the velocity $\overline{\upsilon}_x$ increases with $\theta$ in the steady state. These observations agree with Silbert \textit{et al}~\cite{silbert2001}. Steady state flows are also obtained in this $\theta$ range when the base is vibrated (see Fig.~\ref{fig:steady}); for brevity, the evolution of $\overline{\upsilon}_x$ for only three combinations is shown. 
It is worth noting that the flow may be initiated for the vibrated base at those inclinations where grains do not even mobilize for the fixed base, i.e., below $23^{\circ}$. The objective here is to examine the flow over an inclined vibrated base with respect to the fixed base. Hence, our analysis hereafter will be based on $\theta$ varying between $23^{\circ}$ and $29^{\circ}$. The steady-state velocity for a vibrating scenario is always larger than its fixed-base counterpart for a given $\theta$, as we will see. It is important to note that we consider a loose grain arrangement with an initial volume fraction of $\phi_i \approx 0.22$. We have verified that the resultant flow is independent of the initial packing configuration as shown in Fig.~\ref{fig:phi_i} of appendix \ref{app:phi_i}.

\begin{figure*}
\centering
\includegraphics[scale=0.39]{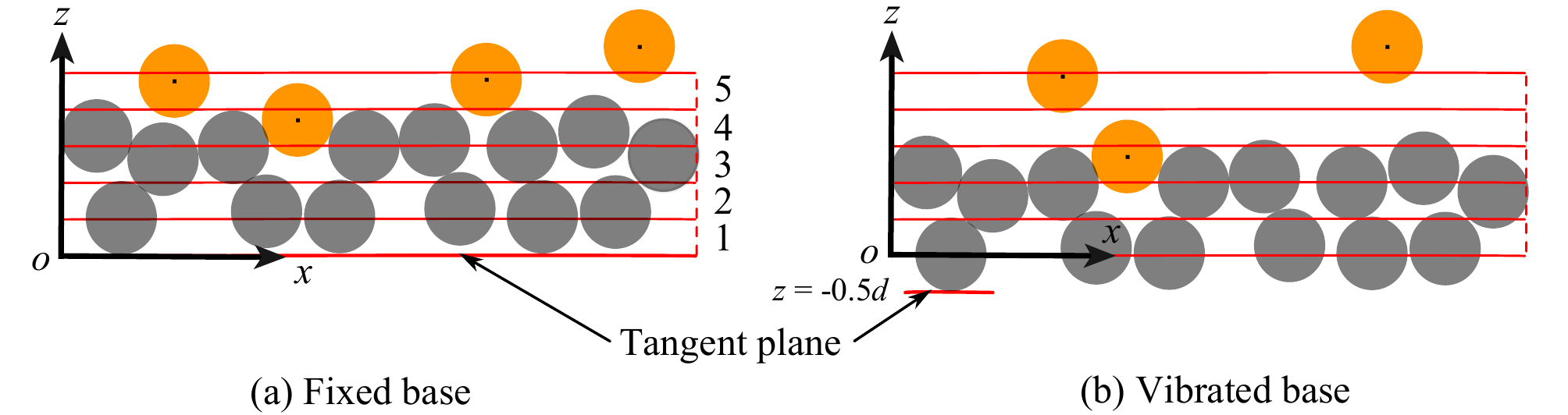}
\caption{\label{fig:zbin} 
(a) A schematic presenting the side view of the fixed base, where $z = 0$ represents the bottom surface of the base.
Gray-colored grains form the bumpy base.
For illustration, five bins are presented in red, which are fixed with reference to the coordinate system $x-y-z$.
No flowing grains are observed in the first three bins.
The center of the lowest moving grain (depicted in orange) lies in the fourth bin centered at $z=1.75 d$.
(b) In the vibrated state, the lower-most position achieved by the bottom surface of the base is $z = -0.5d$. Here, as illustrated, a flowing grain may be found in the third bin at $z=1.25 d$, but the number of such instances are few.}
\end{figure*}

Next, we examine the profiles of velocity $\upsilon_x$, volume fraction $\phi$, and mass flow rate $Q_x$ for the fixed and vibrated bases.
These flow profiles are plotted with reference to the fixed coordinate system $x-y-z$, where $z = 0$ represents the bottom surface of the fixed base. The latter is defined as the plane tangent to the base at its bottom; see Fig. \ref{fig:zbin}(a). Recall that the base has a thickness of approximately two particle diameters. The lowest position of the vibrated base, more precisely the base's bottom surface, varies from $z_b = z = -A~\text{to}~A$. Therefore, for the greatest amplitude $A=0.5d$ considered in this study, the lowest position attained by the bottom surface of the base is $z_b = -0.5d$; see Fig. \ref{fig:zbin}(b). We employ bins of size 0.5$d$ for averaging. Hence, for the fixed base, no moving grains exist in the first three bins, spanning between $z=0$ and $1.5d$ as displayed schematically in Fig. \ref{fig:zbin}(a). The topography of the base accommodates flowing grains in the fourth bin, centered at $z=1.75d$. When the vibrated base moves downward, the third bin centered at $z=1.25d$ may also receive a few grains as depicted in Fig. \ref{fig:zbin}(b), although these are small in number and hence may be ignored. In other words, the data points in the first three bins at the bottom boundary of the flow get omitted. 
Thus, the data are plotted from $z = 1.75d$ onwards in all cases.
Furthermore, we only present those data points where the standard error is less than $2 \%$ of the mean value. Consequently, the data points near the top boundary of the flow get omitted.
\subsection{Flow over fixed base}
\label{sec:fix}
\begin{figure}
\centering
\includegraphics[scale=0.3]{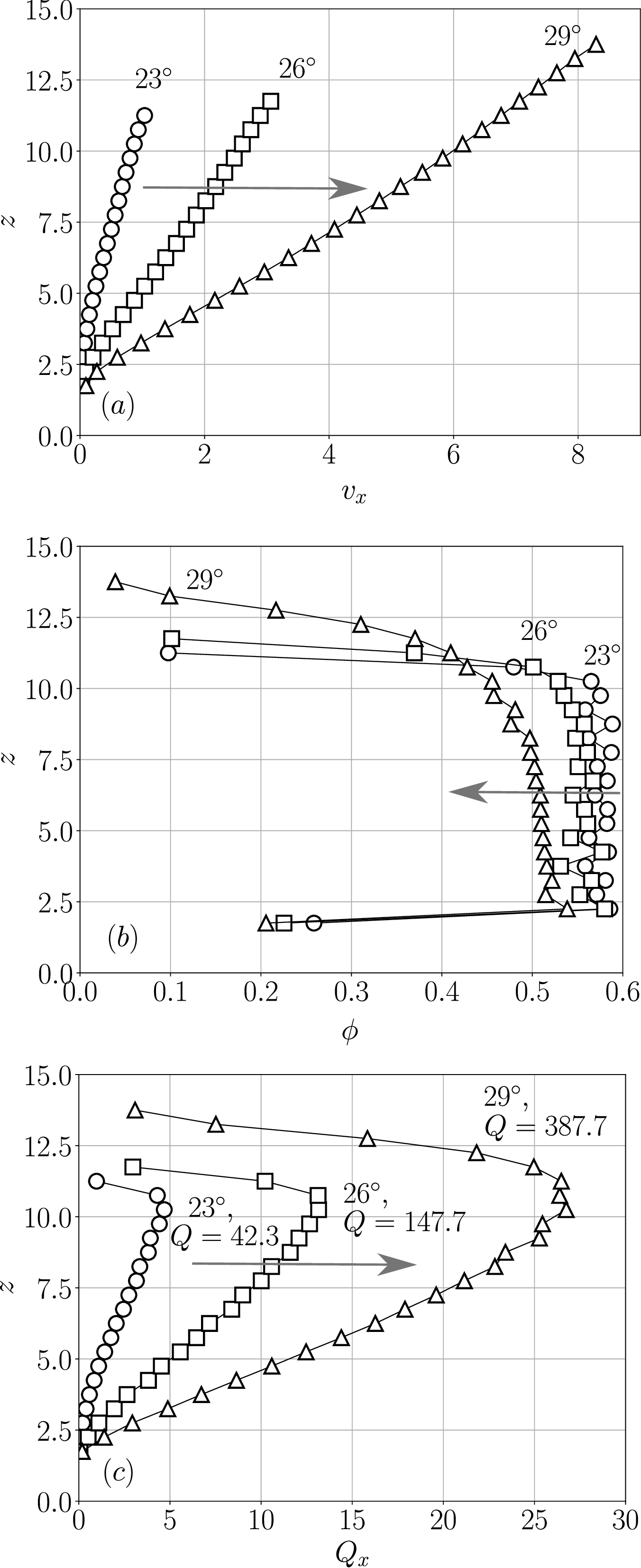}
\caption{Profiles of the (a) flow velocity $\upsilon_x$ {scaled by $\sqrt{gd}$}, (b) volume fraction $\phi$, and (c) mass flow rate $Q_x$ 
{scaled by $\rho d^3 \sqrt{g/d}$} for the fixed base. The arrow direction represents the effect of increasing $\theta$ on the variation of the quantity of interest.}
\label{fig:fixprofiles}
\end{figure}

Figure \ref{fig:fixprofiles} presents the profiles of flow velocity $\upsilon_x$, volume fraction $\phi$, and mass flow rate $Q_x$ for $\theta = 23^{\circ}, 26^{\circ}$ and $29^{\circ}$.
The error bars, corresponding to the standard error, are not shown as these are smaller than the size of markers.
We note in Fig.~\ref{fig:fixprofiles}(a) that the flow velocity rises as $\theta$ increases. For all $\theta$, $\upsilon_x$ increases with $z$, and the profiles $\upsilon_x(z)$ are found to be largely linear, in agreement with what is reported~\cite{silbert2003,gaudel2019} for thin granular systems such as ours. In Fig.~\ref{fig:fixprofiles}(b), we observe that the volume fraction becomes nearly constant for a given $\theta$ in the region away from the base and top. Furthermore, the granular assembly dilates as $\theta$ increases, which is evidenced by decreasing $\phi$ in the bulk. The profiles of mass flow rate $Q_x$ are displayed in Fig.~\ref{fig:fixprofiles}(c). Contrary to $\upsilon_x$, $Q_x$ displays a non-monotonic 
variation with the {depth}, wherein it first increases as we move upwards from the base and then decreases near the top. 
{Variation of $Q_x$ in the top layers is similar to that of $\phi$. 
The sharp change in the volume fraction $\phi$ near the top compared to its nearly constant value in the bulk causes the non-monotonic response in $Q_x$.
Moreover, the rate of change in $\phi$ near the top surface reduces as $\theta$ increases.
}
Furthermore, as expected, the depth-averaged mass flow rate $Q$ monotonically grows as the base becomes steeper; see Fig.~\ref{fig:fixprofiles}(c). Note that this is the maximum $Q$ that may be achieved at a given angle of inclination for the fixed base.
%
%
\subsection{Vibrated base: Role of vibration frequency $f$}
\label{sec:rolef}

\begin{figure}
\centering
\includegraphics[scale=0.3]
{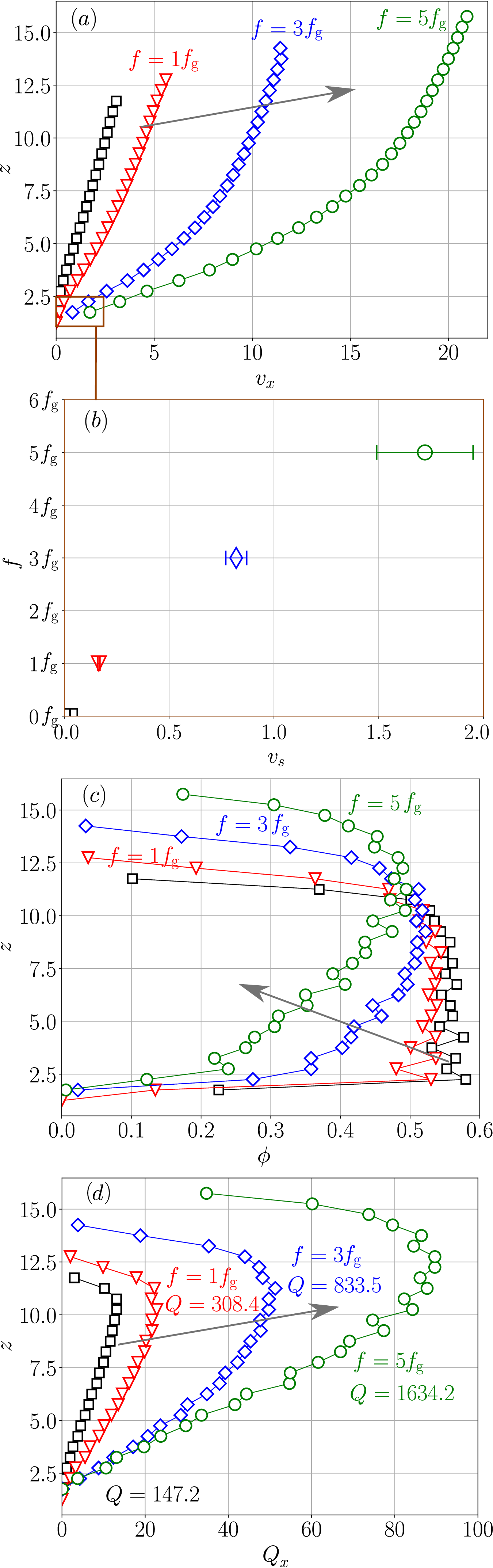}
\caption{(a) Depthwise variation of the flow velocity $\upsilon_x$. 
(b) The slip velocity is $\upsilon_s = \upsilon_x \vert_{z=1.75d}$. The lowest moving layer is located nearly at $z = 1.75d$.
Depthwise variation of the (c) volume fraction $\phi$ and (d) mass flow rate $Q_x$. The arrow direction indicates the effect of increasing frequency on the variation of the quantity of interest. 
The vibrated system has $A = 0.1 \, d$ and $\theta = 26^{\circ}$. Results for the fixed base are in black.} 
\label{fig:vibprofilesf26}
\end{figure}
%
%
\begin{figure}
\centering
\includegraphics[scale=0.3]
{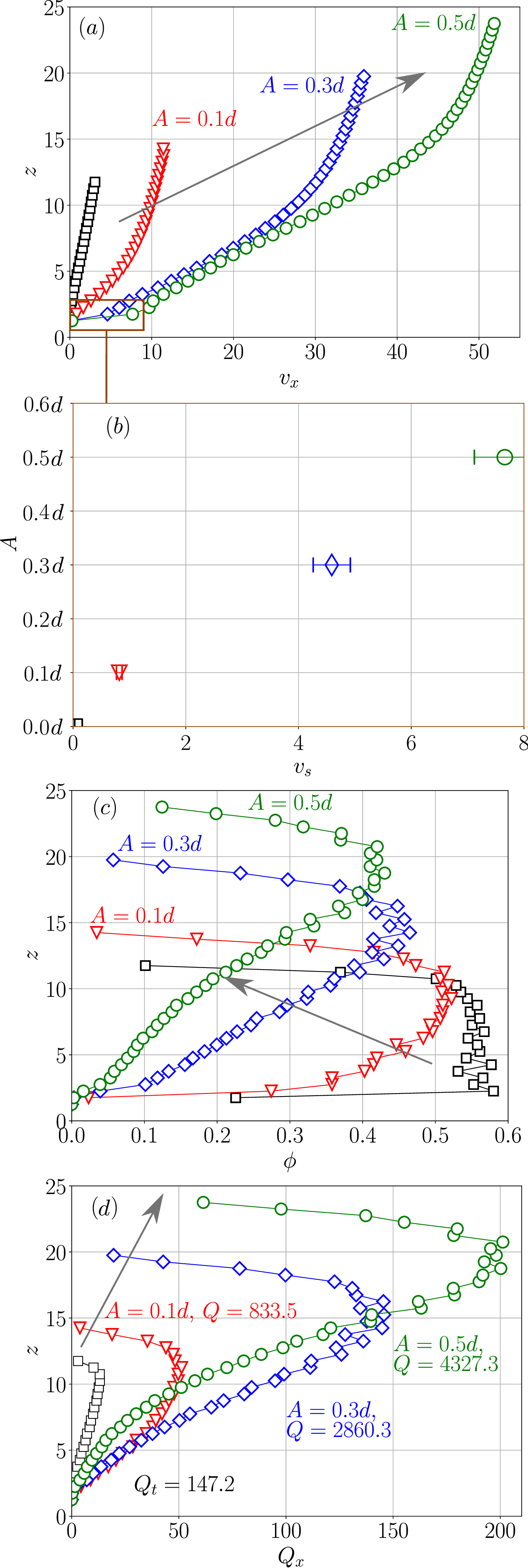}
\caption{(a) Depthwise variation of the flow velocity $\upsilon_x$. 
(b) The slip velocity is $\upsilon_s = \upsilon_x \vert_{z=1.75d}$. The lowest moving layer is located nearly at $z = 1.75d$.
Depthwise variation of the 
(c) volume fraction $\phi$ and (d) mass flow rate $Q_x$. The arrow direction indicates the effect of increasing amplitude on the variation of the quantity of interest.
The vibrated system has $f=3$$f_g$  and $\theta = 26^{\circ}$. Results for the fixed base are in black.}
\label{fig:vibprofilesA26}
\end{figure}
We now examine the scenarios where the base is vibrated sinusoidally with amplitude $A$ and frequency $f$. Figure~\ref{fig:vibprofilesf26} compares the profiles $\upsilon_x$, $\phi$, and $Q_x$ for the fixed and vibrated bases. For the fixed amplitude $A=0.1d$ and inclination angle $\theta = 26^{\circ}$, the profiles are presented for three frequencies. As shown in Fig.~\ref{fig:vibprofilesf26}(a), the velocity profiles for the vibrated base shift towards the right, demonstrating an increase in the flow velocity through the depth. 
We observe that the profiles of $\upsilon_x$ for the vibrated base are non-linear, unlike the fixed base.
In addition, we also observe an increase in the slip velocity at the base with increasing $f$; see Fig.~\ref{fig:vibprofilesf26}(b). 
We define the slip velocity $\upsilon_s = \upsilon_x \vert_{z=1.75d}$, as the lowest mobile layer is located nearly at $z = 1.75d$ due to the base's vibrations as explained above.
The slip velocity was found to be nearly zero for the fixed base even for the highest inclination angle $\theta=29^\circ$; see Fig.~\ref{fig:fixprofiles}(a).
These findings about the augmented slip rate suggest that external shaking dominates the dynamics of flowing grains overcoming the effects of base roughness.
This result is reminiscent of the work of Benedetti \textit{et al.}~\cite{benedetti2012} who investigated the motion of a single angular particle on a transversely vibrated inclined plane. They found that \lq\lq strong friction effects can be overcome by the help of transverse vibration\rq\rq.

Figure~\ref{fig:vibprofilesf26}(c) displays the volume fraction profiles. For all cases, $\phi$ grows as one moves vertically away from the base, achieving a nearly constant value in some regions before reducing further. The depth over which $\phi$ becomes constant shortens as frequency increases. Furthermore, we observe flow dilation with rising $f$, which is expected considering an increase in the external energy input to grains. At higher frequencies, the flow becomes denser in the upper portion, but the volume fraction reduces close to the base. The latter corresponds well to the enhanced flow velocity near the base; see Fig.~\ref{fig:vibprofilesf26}(a).

In Fig.~\ref{fig:vibprofilesf26}(d), we notice a non-monotonic variation of $Q_x$ with $z$, attaining a nearly constant value in a small region near the top for all cases with vibration. It is worth noting that when we vary $f$ from $1$$f_g$ to $5$$f_g$ for $A=0.1~d$, the increase in the depth-averaged flow rate $Q$ is more than $2$ to $10$ times the flow rate on the fixed base, respectively. This finding has major implications for rapid grain transport through inclined chutes.

%
%
\subsection{Vibrated base: Role of vibration amplitude $A$}
\label{sec:roleA}
\begin{figure*}
\centering
\includegraphics[scale=0.33]
{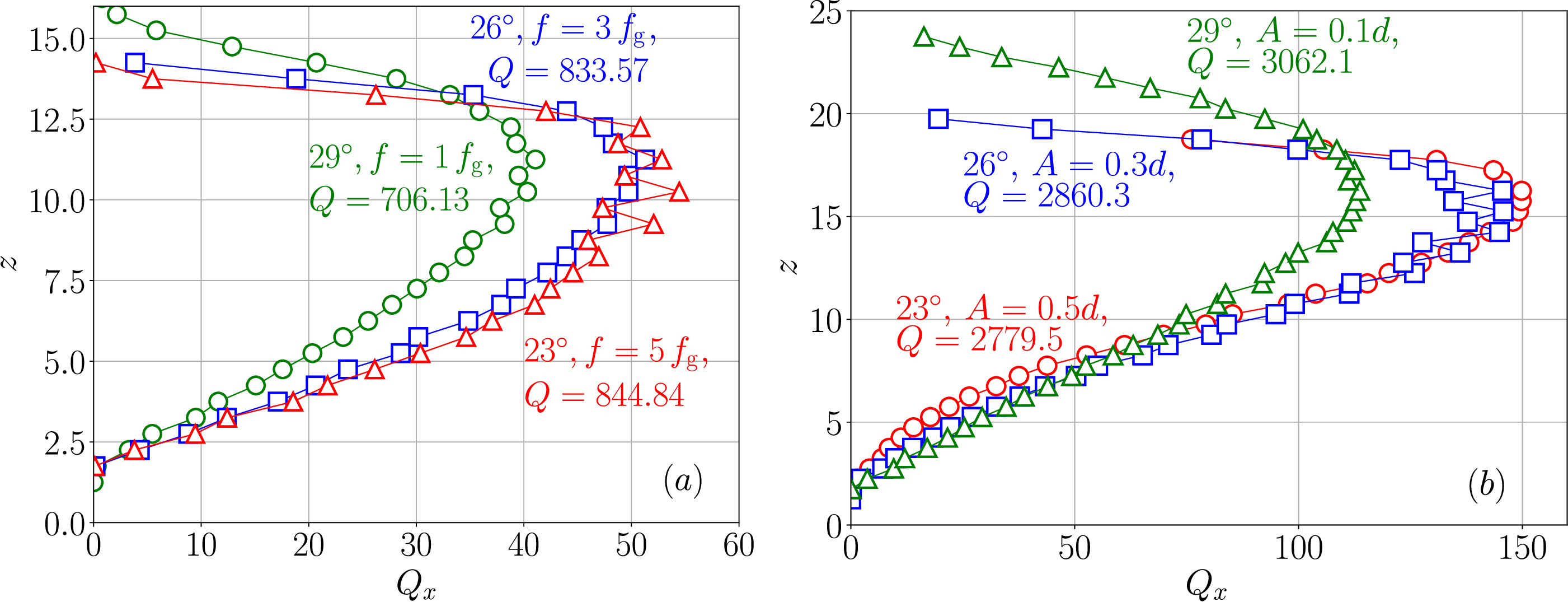}
\caption{Profiles of the mass flow rate $Q_x$ for the vibrated base at (a) constant $A = 0.1d$ and various combinations of $\theta$ and $f$ (b) constant $f = 3$$f_g$ and various combinations of $\theta$ and $A$.}
\label{fig:vibQxprofilesAf}
\end{figure*}
Next, we explore the effect of vibration amplitude on $\upsilon_x$, $\phi$ and $Q_x$ in Fig.~\ref{fig:vibprofilesA26}, keeping $\theta$ and $f$ fixed at $26^{\circ}$ and $3$$f_g$, respectively. Note that these profiles are similar to their counterparts for the frequency variation at a constant amplitude in Fig.~\ref{fig:vibprofilesf26}. 
Again, as observed earlier, a finite slip velocity occurs at the base and increases monotonically with the amplitude $A$; see Fig.~\ref{fig:vibprofilesA26}(b).
As shown in Fig.~\ref{fig:vibprofilesA26}(a), the profiles $\upsilon_x$ for the vibrated base displace towards the right with growing $A$, indicating a rise in flow velocity along $z$. 
However, in the lower portions of the flow, the change in velocity is minimal as $A$ increases beyond $0.3d$.  
In the same region, interestingly, the volume fraction reduces considerably with increasing $A$; see Fig.~\ref{fig:vibprofilesA26}(c). 
As previously observed in Fig.~\ref{fig:vibprofilesf26}(c), the depth over which $\phi$ remains constant also decreases with increasing amplitude at a constant frequency. Moreover, constant packing fraction is observed to occur close to the free surface.
The mass flow rate profiles $Q_x$ are presented in Fig.~\ref{fig:vibprofilesA26}(d), demonstrating a non-monotonic variation with the {depth}. Looking at the profiles for $A=0.3d$ and $0.5d$, the flow rate in the bottom region of the flow is more for the former than the latter, which is expected given the larger volume fraction for $0.3d$. The depth-averaged flow rate $Q$, nonetheless, exhibits a monotonic trend with the amplitude. 
Additionally, for the maximum $A$ employed at $f=3 f_g$, the flow thickness doubles and the corresponding $Q$ increases roughly by $30$ times the value for corresponding flows over the fixed base at the same inclination.
Before we proceed further, we mention in passing that the velocity profiles shown in Figs.~\ref{fig:vibprofilesf26}(a) and \ref{fig:vibprofilesA26}(a) generally do not follow the Bagnold curve; see appendix \ref{app:bagnold}. This is because a vibrating base modifies the grain density and motion in its neighborhood.
\subsection{Depth-averaged mass flow rate $Q$}
\label{sec:Q}
We have examined so far the effect of amplitude and frequency separately on the flow rate for a fixed inclination angle. Now, keeping the amplitude constant at $A=0.1d$, Fig.~\ref{fig:vibQxprofilesAf}(a) compares the profiles of the flow rate for three combinations of $f$ and $\theta$ given in the format [$\theta,f$]: [$23^\circ,5$$f_g$], [$26^\circ,3$$f_g$], and [$29^\circ,1$$f_g$]. Interestingly, the profiles representing the first two cases largely overlap, leading to nearly the same depth-averaged mass flow rate $Q$. Similarly, Fig.~\ref{fig:vibQxprofilesAf}(b) reports approximately the same mass flow rate $Q$ for two combinations of $A$ and $\theta$ when the system is vibrated at a constant frequency of $f=3$$f_g$. These findings indicate that a nearly identical mass flow rate can be obtained for the vibrated base using an appropriate combination of $\theta$, $A$, and $f$. 
Naturally, the next step is to look for a parameter involving $A$ and $f$ for a given $\theta$, which controls the mass flow rate $Q$. To this end, we carry out a scaling analysis, which we discuss next.
%
%
%
\subsection{Scaling analysis for $Q$}
\label{sec:scale}
\begin{figure}
\centering
\includegraphics[scale=0.31]{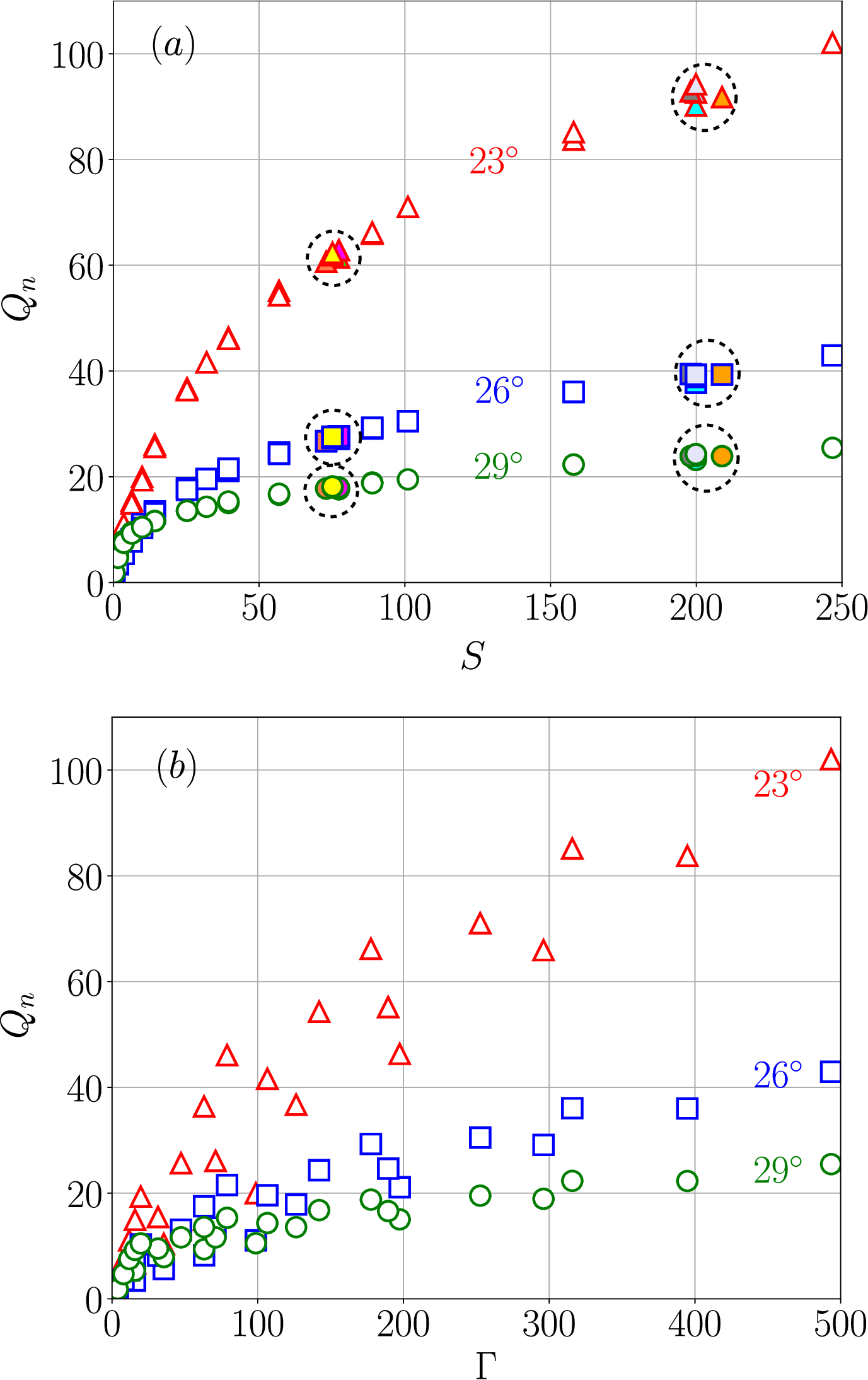}
\caption{\label{fig:QSGamma} Variation of the scaled flow rate $Q_n$ with (a) $S$ and (b) $\Gamma$ for $\theta = 23^{\circ}$, $26^{\circ}$, and $29^{\circ}$. The dashed circles represent clusters of data points having similar $S$ values. Here, $A$ and $f$ ranges from $0.1d-0.5d$ and $1-5~$Hz, respectively. Other parameters $e, \mu$, and $\phi_i$ are kept constant.}
\end{figure}
The mass flow rate $Q$ may be expressed through the functional dependence
\begin{equation}
Q = F(\rho,g,d,A, f, h_0, e,\mu,\theta,\phi_i),
\label{eq:Q0}
\end{equation}
which may be non-dimensionalized as
\begin{equation}
\overline{Q} := \frac{Q}{\rho d^3 \,\sqrt{g/d}} = \overline{F} \left(\frac{A^2 \omega^2}{g d}, \frac{A \omega^2}{g}, \frac{h_0}{d}, e, \mu, \theta, \phi_i \right),
\label{QfunND}
\end{equation}
where $\omega = 2 \pi f$ is the angular frequency. The dimensionless numbers $S = A^2 \omega^2 / gd$ and $\Gamma = A \omega^2 / g$ arise naturally. We note that $A/d$ is not an independent parameter, and is given by the ratio $S/\Gamma$. 
Now, as mentioned earlier, the system dynamics is unaffected by the initial packing fraction $\phi_i$; see Fig.~\ref{fig:phi_i}. Also, we keep $e$ and $\mu$ constant in this study. Accordingly, we reduce (\ref{QfunND}) to 
\begin{equation}
\overline{Q} = \overline{F} \left(S, \Gamma, \frac{h_0}{d}, \theta \right).
\label{Qfinal}
\end{equation}
Note that $S$ and $\Gamma$ signify the energy and inertia of the system, respectively, and are reported to govern different phases of the flows~\cite{eshuis2007}. Thus, we explore both these parameters for various inclinations and base vibrating conditions.
The flow rate of our vibrated inclined granular flows is now characterized utilizing such dimensionless numbers. To this end, for a given $\theta$, we define $Q_n ={Q}_{\upsilon}/{Q}_{f}$, where ${Q}_{\upsilon}$ and ${Q}_{f}$ are scaled mass flow rates over vibrated and fixed bases, respectively.

%
We now examine the variation of $Q_n$ with $S$ and $\Gamma$ for three values of $\theta$ in Fig.~\ref{fig:QSGamma}. The data for each $\theta$ corresponds to $35$ combinations of the amplitude $A$ and frequency $f$, resulting in $S$ and $\Gamma$ lying in the ranges, respectively, $[0.39,246.74]$ and $[3.95,493.48]$. Remarkably, as Fig.~\ref{fig:QSGamma}(a) displays, $Q_n$ values for different $S$ lie on a single curve for a given $\theta$. However, such a scaling is not obtained for $\Gamma$ in Fig.~\ref{fig:QSGamma}(b). 
It is worth remarking that the scatter in data shown in Fig.~\ref{fig:QSGamma}(b) appear significant at small angles, i.e. $\theta=23^{\circ}$. 
Nonetheless, this is true for all the inclinations, which is not presented for brevity here.
Figure~\ref{fig:QSGamma}(a) also shows that $Q_n$ grows as $\theta$ decreases for a given $S$, indicating that the vibrated base significantly influences flows, especially at lower inclination angles. In other words, gravity dominates $S$ at steeper angles. These findings also indicate that the product of $A$ and $\omega$ controls the flow rather than $A$ and $\omega$ individually, which is in concurrence with the predictions of \cite{soto2004}.
It is also worth noting that vibrating the base at $23^{\circ}$ inclination increases the depth-averaged mass flow rate by $100$ times compared to the fixed base. This enhancement is significantly larger than at higher inclinations, wherein $Q$ rises by 40 and 25 times at $26^{\circ}$ and $29^{\circ}$, respectively.

To further confirm the collapse of the data with $S$, a few more combinations of $A$ and $f$ are considered so that $S \approx 77$ and $200$. When this is done, some of the amplitude and frequency values exceeded their earlier range given in Sec.~II.
For the above two $S$ values, the scaled flow rate $Q_n$ lies on the original curve corresponding to a given $\theta$, as displayed by a cluster of data points in the dashed circles in Fig.~\ref{fig:QSGamma}(a). This indicates the universality of the parameter $S$ to characterize the vibrated inclined flows.

%
Before we conclude, it is pertinent to revisit the key observations of previous studies.
Eshuis \textit{et al}.~\cite{eshuis2007} considered horizontal vibrated beds and clearly distinguished the regimes governed primarily by $S$ and $\Gamma$.
Our study, however, focuses on inclined granular flows, aiming to understand the possibility of regulating the flow rate through base vibrations. 
Other works on inclined granular flows with transversely vibrated bases primarily employed $\Gamma$ to interpret results \cite{gaudel2016, gaudel2019, ambrosio2023}, without considering the dimensionless number $S$ that we found above to be more relevant than $\Gamma$.
Zhu \textit{et al}.~\cite{zhu2020} employed a system like ours, investigating earthquake-induced landslides by simulating seismic activity at low inclinations. They found that normal base vibrations with a constant acceleration amplitude of $A \omega^2 = 0.5 g$ increased the mobility of grains, leading to longer landslide runouts.
However, no attempt was made to understand this observation in terms of either $S$ or $\Gamma$. Nevertheless, their observations are consistent with our findings above that base vibrations may greatly augment the mass flow.
%

\section{Conclusion}
\label{sec:con}
We investigated the effect of normal base vibrations on inclined granular flows by utilizing discrete element method. We found an appreciable change in the flow profiles when the base is vibrated, with an overall increase in the flow velocity and dilation across the depth. As a consequence, the depth-averaged mass flow rate $Q$ was greatly enhanced when the base was vibrated; $Q$ increased roughly by $100$ times for $23^{\circ}$, up to $40$ times for $26^{\circ}$ and $25$ times for $29^{\circ}$ in comparison to
corresponding flows over a fixed base at the same inclinations. This will have important industrial applications in the design of high-throughput conveyor systems. At the same time, it was possible to keep $Q$ nearly constant by appropriately adjusting the base’s inclination $\theta$, and its vibrational amplitude $A$ and frequency $f$. {This, in turn, will have important implications for the controllability of such flows}. We also observed that the velocity profiles at all base inclinations considered here exhibited a non-linear variation with the flow depth, unlike the predominantly linear trend reported for fixed base flows.
Finally, utilizing dimensional analysis, we demonstrated that the flow rate for the vibrated base, scaled by its fixed-base counterpart, could not be characterized by the scaled acceleration $\Gamma$, but could be done so in terms of the non-dimensional parameter $S$. 
This suggests that revisiting research on vibrated granular flows with $S$ as the primary control parameter may yield valuable insights.
%
%
\appendix

\setcounter{figure}{0}
\renewcommand{\thefigure}{A\arabic{figure}}

%
%
%
%
\section{Effect of initial volume fraction}
\label{app:phi_i}
\begin{figure}
\centering
\includegraphics[scale=0.33]{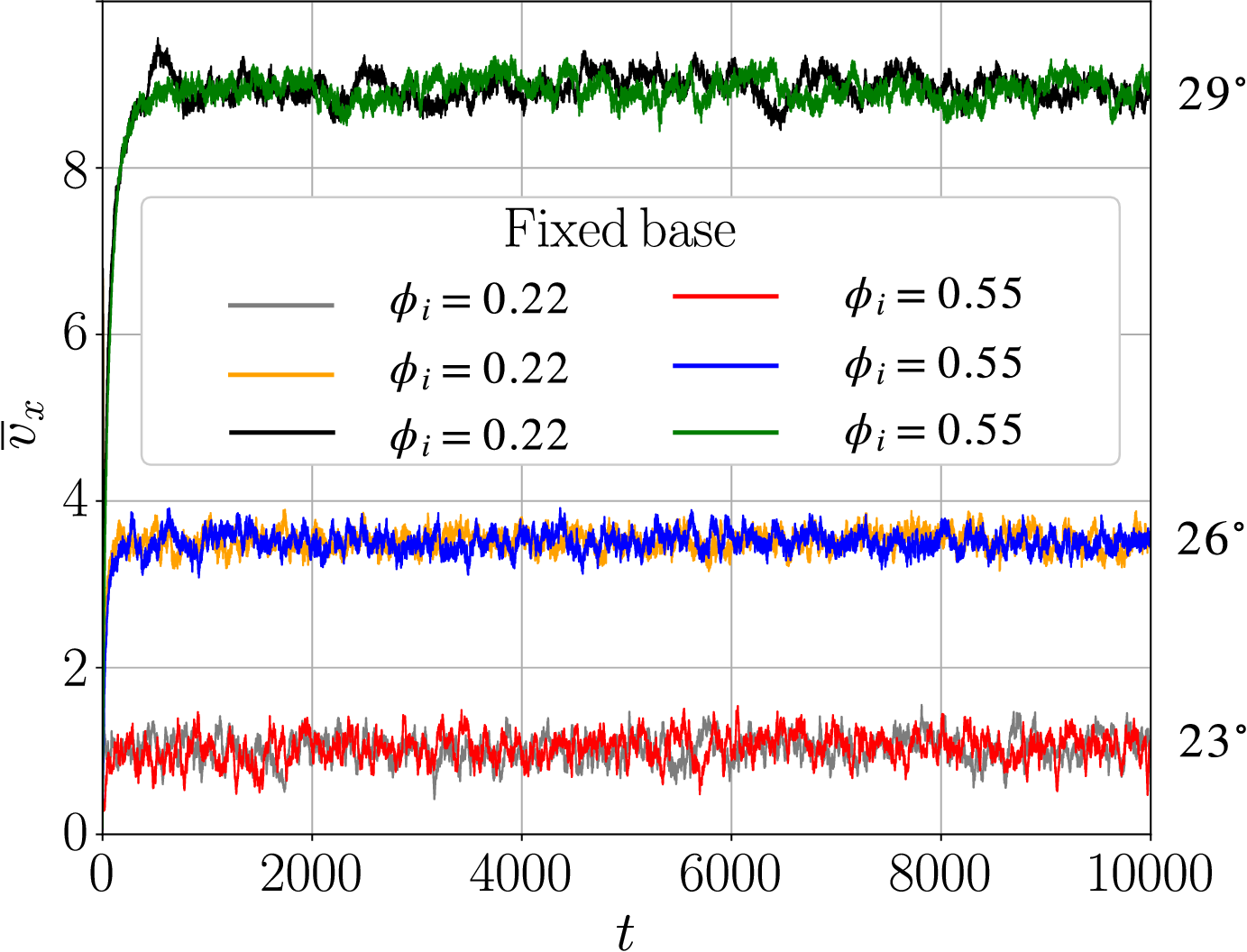}
\caption{\label{fig:phi_i} Temporal evolution of the average translational velocity $\overline{\upsilon}_x$ on the fixed base for two initial volume fractions $\phi_i$ at three different inclination angles. The flow velocity is scaled by $\sqrt{gd}$.}
\end{figure}

The initial volume fraction $\phi_i \approx 0.22$, indicating a loose arrangement of grains. The results presented in this work are independent of the initial configuration, which is confirmed by considering a denser initial packing corresponding to the volume fraction of $\phi_i \approx 0.55$. Figure~\ref{fig:phi_i} displays the temporal evolution of the average translational velocity, $\overline{\upsilon}_x(t)$, of the flow over the fixed base for three different inclination angles. It is evident that the steady state is largely unaffected by the initial preparation, suggesting that $\phi_i$ does not affect the steady flow conditions of the system under investigation.\\

\section{Velocity profiles}
\label{app:bagnold}
\begin{figure*}
\centering
\includegraphics[scale=0.3]
{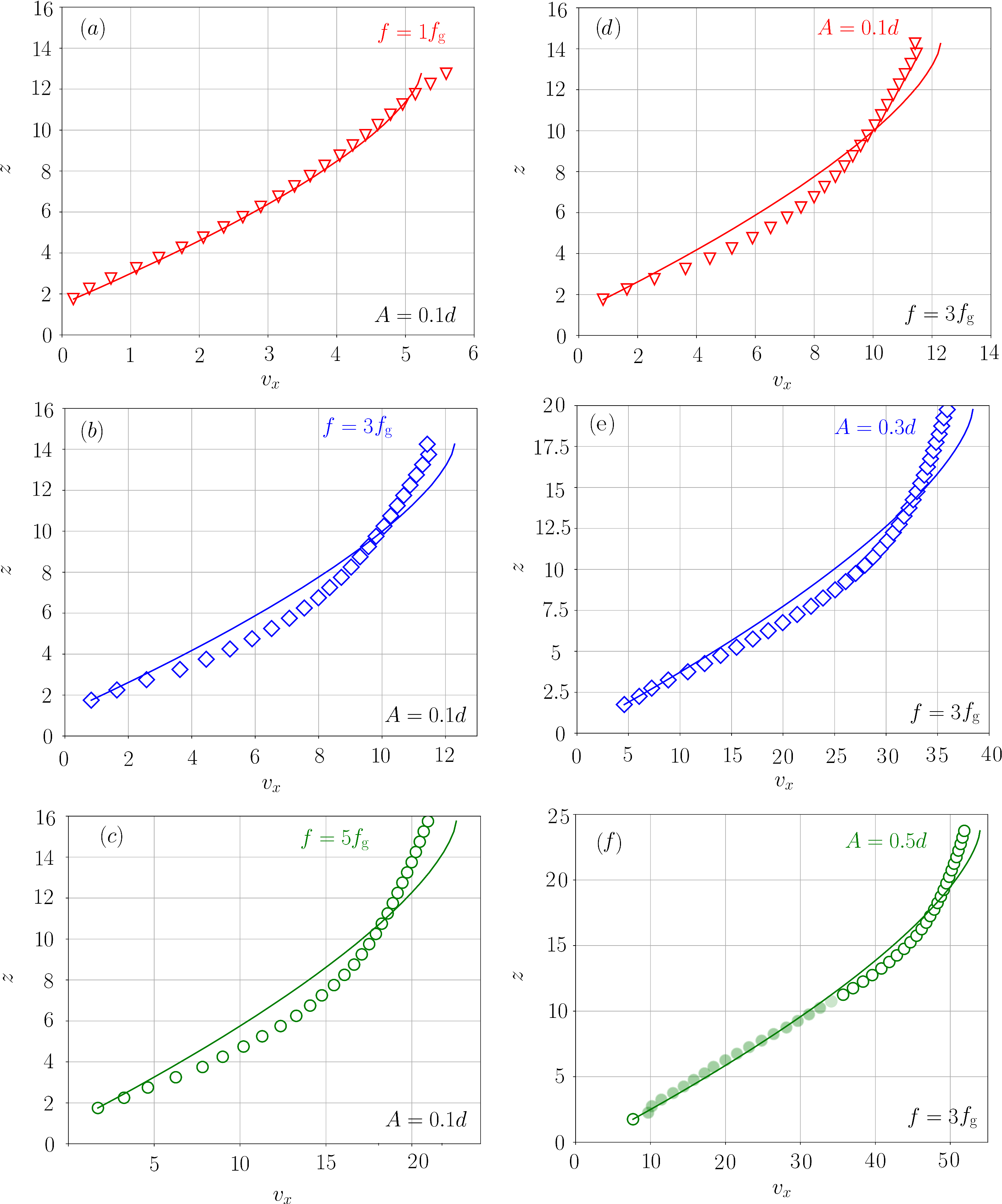}
\caption{\label{fig:bagnold} The depthwise variation of the flow velocity $\upsilon_x$ compared with generalized Bagnold profiles (solid lines) represented by Eq.~(\ref{eq:genbagnold}). 
[(a)-(c)] The frequency varies at a constant amplitude $A=0.1 \, d$.
[(d)-(f)] The amplitude changes at a fixed frequency $f=3f_g$. The base is inclined at $\theta = 26^{\circ}$.}
\end{figure*}

\noindent The Bagnold curve represented by the expression below is fitted to the velocity profiles obtained from simulations
\begin{equation}
\label{eq:bagnold}
\upsilon_x(z) = A_{b} ~ H^{3/2} ~ \bigg[1-\bigg(\frac{H-z}{H}\bigg)^{3/2}\bigg],
\end{equation}
where $H$ is the effective flow height and $A_{b}$ is a fitting parameter~\cite{silbert2001}.
Equation~(\ref{eq:bagnold}) assumes no slip velocity at the base, i.e. $\upsilon_x = 0$ at $z = 1.75d$, and applies only to flows with a constant density. 
Thus, the Bagnold profile is obtained for the bulk region, away from the top and bottom boundaries~\cite{silbert2001}. The Bagnold profile given by Eq.~(\ref{eq:bagnold}) is generalized by considering the basal slip velocity $\upsilon_{x0}$  \cite{hill2022}
\begin{equation}
\label{eq:genbagnold}
\upsilon_x(z) = \upsilon_{x0} + A_{b} ~ H^{3/2} ~ \bigg[1-\bigg(\frac{H-z}{H}\bigg)^{3/2}\bigg].
\end{equation}

We use the velocity data from Figs.~4(a) and 5(a) to generate Figs.~\ref{fig:bagnold}(a-c) and Figs.~\ref{fig:bagnold}(d-f), respectively. As shown in Fig.~\ref{fig:bagnold}(a), for the minimum values of $A = 0.1 d$ and $f = 1 \, f_g$, the data follows the generalized Bagnold profile, except near the top surface. However, as $A$ and $f$ increase, the velocity profiles deviate from the generalized Bagnold profile; see Figs.~\ref{fig:bagnold}(b-f). In particular, Fig.~\ref{fig:bagnold}(f) displays a gentle concave profile near the base (highlighted by filled symbols), likely caused by a reduced volume fraction, indicating a further deviation from the generalized Bagnold profile at higher base vibrations.\\\\

\bibliography{bib_prf_vib_Sept2024}

\end{document}